\documentclass[preprint,showpacs,preprintnumbers,amsmath,amssymb]{revtex4-2}
\usepackage{graphicx}
\usepackage{bm}
\usepackage{color}
\usepackage{amssymb}
\usepackage{epsfig}
\DeclareGraphicsExtensions{.png,.pdf}

\begin{document}

\title{Non-Hermiticity and Universality}
\author{Pragya Shukla$^{*}$}
\affiliation{ Department of Physics, Indian Institute of Technology, Kharagpur-721302, West Bengal, India \\\\
$^{*}$ E-Mail: shukla@phy.iitkgp.ac.in}
\date{\today}

\widetext

\begin{abstract}

	We study the statistical properties of the eigenvalues 
of non-Hermitian operators associated with the dissipative 
complex systems.  By considering the Gaussian ensembles of such 
operators, a hierarchical relation between the correlators is obtained. 
Further the eigenvalues are found to behave like 
particles moving on a complex plane under 2-body (inverse square) 
and 3-body interactions and there seems to underlie a deep 
connection and universality in the spectral behaviour of different 
complex systems.
\end{abstract}

\maketitle

..

	The random non-Hermitian operators play a significant role in
the dynamics of variety of complex systems e.g. dissipative quantum
systems \cite{fh1,fz}, chaotic quantum scattering \cite{fy1},
 neural network dynamics \cite{sc},
 statistical mechanics of flux lines
in superconductors with columnar disorder \cite{fte},
classical diffusion in random
media \cite{cw}, biological growth problems \cite{ns}.
A detailed knowledge of the statistical
properties of their eigenvalues and eigenvectors therefore is very 
 desirable. However, so far, the information is available only 
for a few specific 
cases e.g \cite{fh1,fz,cb,fy2}.

	The  complicated 
nature of the interactions (or a lack of detailed information about 
them) in a system introduces a degree of randomness in the matrix 
representation of the associated 
operators. The indeterminacy of the interactions permits each matrix element 
to be described only by a distribution 
of possible values, resulting in a random matrix. However the 
choice of a suitable random matrix model for an operator of a 
complex system is very sensitive to the nature of its complexity. The 
statistical spectral analysis of different complex systems requires, 
therefore, a thorough probing of a wide range of random matrix ensembles  
which is not an easy task. It is highly desirable, if possible, to find a 
common mathematical formulation for various systems 
where the information about the system enters through a parameter. The 
possibility of such a formulation has already been shown for the Hermitian 
operators \cite{ps1}. This encourages us to seek the same for non-Hermitian 
operators too. 
We explore the non-Hermitian random matrix ensembles (NHRE) 
with a Gaussian distribution and obtain the 
 formulation using exact analytical methods. 
This is achieved by showing that the eigenvalue distributions  
of various NHREs appear as non-equilibrium stages of a Brownian type 
diffusion process. Here the eigenvalues evolve with respect to a 
parameter related to the complexity of the system associated with the NHRE.
The solution of the diffusion equation  
for a given value of the parameter gives, therefore, the  
distribution of the eigenvalues for the corresponding system. Using the  
 diffusion equation, we also obtain the hierarchic relations
among eigenvalue correlations.  
 
	The connection of various NHREs through a diffusion process can 
further be used to reveal a very interesting feature about their 
eigenvalue dynamics in the following way. A suitable transformation 
can reduce the diffusion equation to the Schrodinger equation of a 
classically integrable Hamiltonian, thereby mapping the eigenvalue 
distribution of a 
general NHRE to a non-stationary state of the Hamiltonian.
The latter,
a variant of Calogero-Sutherland (CS) Hamiltonian in two dimension,
is a generator of the dynamics of $N$ particles interacting via long-ranged
two body and three body interactions and confined by a harmonic oscillator
potential \cite{cam,ak}. The 
information about the spectral correlators of a given NH system can therefore 
be obtained  from the particle correlations of the CS system. 
	As well-known, the CS Hamiltonian is an integrable
system 
with particles evolving in an ordered way with respect to time;
 this indicates a strong correlation between various particle
states at different times. Our mapping thus implies that the 
eigenvalues evolve in a highly ordered, correlated way as the degree or the
nature of the complexity changes. This would also indicate a strong 
correlation  
between the statistical nature of the eigenvalues of  different 
complex systems. 
Note another study has already indicated the universality in the 
eigenvalue statistics of 
the operators in the regime of weak non-Hermiticity \cite{fy3}.

We consider an ensemble of $N\times N$ non-Hermitian matrices $H$
defined by a Gaussian measure
$\tilde\rho (H,y,x)$ where
 $\tilde \rho  \propto {\rm exp}[
- \sum_{s=1}^\beta \sum_{k, l}
 (y_{kl;s} H_{kl;s}^2 + x_{kl;s} H_{kl;s} H_{lk;s})] = C \rho(H,y,x) $
 with $C$ as the normalization constant,
 $y$ and $x$ as sets of the variances and covariances of various
matrix elements.
  Here the subscript $s$ on a variable refers to  one of its
components, i.e real ($s=1$) or imaginary ($s=2$) part, with $\beta$ as
 total number of the components.
The above choice of $\rho$ is made so as to include a large class of
NHRE.

	A non-Hermitian matrix can be diagonalized by a transformation of
the type $\Lambda = U H V$ with $\Lambda$ as the matrix of eigenvalues
 $\lambda_j $ and $U$ and $V$ as the left and
right eigenvector matrices
respectively. Let us first consider the case of an ensemble of
 non-Hermitian complex matrices ($\beta=2$). Here the eigenvalues
$\lambda_j \equiv \sum_{r=1}^2 (i)^{r-1} \lambda_{jr} $, in general,
($r$ referring to the components of the eigenvalues)
are distributed over an area in the complex plane.
Let  $\tilde P(z,y,x)$ be the probability of finding eigenvalues
$\lambda_i $  of $H$ between $z_i$ and $z_i+{\rm d}z_i$ for  
 given sets $y$ and $x$,
\begin{eqnarray}
\tilde P(z,y,x)=  C\int f(z,z^*)\rho (H,y,x){\rm d}H 
\end{eqnarray}
with $z \equiv \{z_i\}$ and 
$f(z,z^*) = \prod_{i=1}^{N}\delta(z_{i}-\lambda_{i})
\delta(z_i^*-\lambda_i^*)$.

 	The degree of difficulty associated with solving the integral
eq.(1) motivates us to seek another route. We attempt
to obtain an evolution equation, in a well-known form, for $P = \tilde P/C$  
due to changing
distribution parameters. (The prior familiarity with the equation 
can then be used to obtain the information about its solution $ P$).   
For this pupose, we  consider a combination of the parametric derivatives 
of $P$, namely the sum $S$,
$S\equiv   \sum_{s=1}^{\beta}  \sum_{k,l}
 \left[ A_{kl;s} {\partial P \over\partial y_{lk;s}}
 +  B_{kl;s}  {\partial P \over\partial x_{kl;s}}\right]$
and express it in terms of the eigenvalue derivatives 
of $P$. Here 
$A_{kl;s} =   y_{kl;s} [\gamma + 2 (-1)^s x_{lk;s} ]$ and
$B_{kl;s} = [\gamma x_{kl;s} + (-1)^s x_{kl;s} x_{lk;s}
+ (-1)^s y_{kl;s} y_{lk;s}] $
(with $\gamma$ as an arbitrary parameter and, as shown later,  
 its arbitrary value gives the freedom to choose the end of the evolution). 
This would require a knowledge of the rates of change of the 
eigenvalues as well as
the eigenvectors due to a small change in the matrix element $H_{kl}$,
given as follows,

\begin{eqnarray}
{\partial \lambda_n \over \partial H_{kl;s}} =
   i^{s-1} U_{nk} V_{ln};
\sum_{k,l;s} {\partial \lambda_n \over\partial H_{kl;s}} H_{kl;s}
 = \lambda_n \nonumber \\ 
 \sum_{k,l;s} (-1)^{s-1}{\partial \lambda_n \over\partial H_{kl;s}}
  {\partial \lambda_m \over\partial H_{lk;s}}  = \beta \delta_{mn} 
\nonumber \\
 \sum_{k,l;s} (-1)^{s-1}
{\partial^2 \lambda_n \over\partial H_{kl;s} H_{lk;s}}
= \sum_{m} {2\beta \over \lambda_n - \lambda_m} \nonumber \\
{\partial U_{nr} \over\partial H_{kl;s}} =
 \sum_{m\not=n}
{U_{nk}\over {\lambda_n -\lambda_m}}
{\partial \lambda_m \over \partial H_{rl;s}},\nonumber \\
{\partial V_{rn} \over\partial H_{kl;s}} =
 \sum_{m\not=n}{V_{ln}\over {\lambda_n -\lambda_m}}
{\partial \lambda_m \over \partial H_{kr;s}} 
\end{eqnarray}

The parametric-dependence of $P$
in sum $S$ enters only through $\rho$ and
as ${\partial \rho \over\partial y_{kl;s}}
= - H_{kl;s}^2 \rho$,
 ${\partial \rho \over\partial x_{kl;s}}
=- H_{kl;s}  H_{lk;s} \rho$,
${\partial \rho \over\partial H_{kl;s}} =
-2(y_{kl;s}H_{kl;s}+x_{kl;s} H_{lk;s})$ with
$ {\partial f \over \partial H_{kl;s}}
=-2\sum_r {\partial \lambda_{nr} \over \partial H_{kl;s}}
{\partial f \over \partial z_{nr}}$,
the sum $S$ can be rewritten:  
\begin{eqnarray}
S= \sum_{s=1}^\beta \sum_{k,l}\left[\gamma +
(-1)^s  x_{kl;s} \right] I_{kl;s} + G - C_1 P
\end{eqnarray}
where
 $I_{kl;s}=\sum_{r,n} {\partial \over \partial z_{nr}}
\int f {\partial \lambda_{nr} \over \partial H_{kl;s}}
H_{kl;s}\; \rho \;{\rm d}H$,
$G=\sum_{k,l;s} (-1)^s 
\left[  y_{kl;s} y_{lk;s} {\partial P \over\partial x_{kl;s}}
+  x_{kl;s} y_{lk;s} {\partial P \over\partial y_{lk;s}}   \right] $,
$C_1= (1/ 2) \sum_{k,l;s}
\left(\gamma + (-1)^s x_{kl;s}\right)$.
Further, by using the eqs.(2), one can show that     
$\sum_{k,l;s}  I_{kl;s} =
 \sum_{n,r} {\partial \over \partial z_{nr}} \left(z_{nr} P\right)$
and 
$\sum_{k,l;s}(-1)^s x_{kl;s}  I_{kl;s}
=  \sum_{n,r} \, (-1)^r \left({\partial^2 P\over \partial z_{nr}^2}
- 2 {\partial \over \partial z_{nr}}
\left( {\partial {\rm ln} |\Delta (z)| \over \partial z_{nr}} P\right)\right) - G$;
here  $\Delta_N(z)=\prod_{j<k}^N(z_j-z_k)$.  
A substitution of these equalities in eq.(3), followed by  a comparison of  
the so-obtained form of $S$ with its original definition,    
 gives now a relation between the parametric and eigenvalue derivatives 
of $P$,  
\begin{eqnarray}
 \sum_{s=1}^{\beta}  \sum_{k,l} &&
 \left[ A_{kl;s} {\partial P \over\partial y_{lk;s}}
 +  B_{kl;s}  {\partial P \over\partial x_{kl;s}}\right] + C_1 P = \nonumber \\
&=& \sum_{r=1}^2 \sum_{n=1}^N   
{\partial \over \partial z_{nr}}\left[  (-1)^r 
 {\partial \over \partial z_{nr}}  -   (-1)^r  \, \beta \, 
{\partial {\rm ln} |\Delta_N (z)| \over \partial z_{nr}}
 + \gamma z_{nr} \right] P
\end{eqnarray}

However it is possible to define a parameter $Y$, a function of 
all $y_{kl;s}$ and $x_{kl;s}$ such that the sum on the left hand side of the 
above equation can be reduced to a derivative of $P$ with 
respect to a single parameter $Y$, with $P(z,y,x)=P(z,Y)$. 
As obvious $Y$  should satisfy the condition that
\begin{eqnarray}
 \sum_{s=1}^{\beta}  \sum_{k,l}
 \left[ A_{kl;s} {\partial P \over\partial y_{lk;s}}
 +  B_{kl;s}  {\partial P \over\partial x_{kl;s}}\right]
 = {\partial P\over\partial Y}. 
\end{eqnarray}
As 
${\partial P\over\partial r}=
{\partial P\over \partial Y}{\partial Y\over \partial r}$,  
$r$ taken from sets $y$ or $x$, the above condition can be rewritten 
as $ \sum_{s=1}^{\beta}  \sum_{k,l}
 \left[ A_{kl;s} {\partial Y \over\partial y_{lk;s}}
 +  B_{kl;s}  {\partial Y \over\partial x_{kl;s}}\right] = 1$.
The form of $Y$, fulfilling the condition, can
therefore be obtained by solving following equations \cite{ps1}
(for all $k,l$ and $s$ values):
${{\rm d}y_{kl;s} \over A_{kl;s}}
= {{\rm d}x_{kl;s} \over B_{kl;s}}
={{\rm d}Y \over 1}$.
The solution $Y$ turns out to be $Y=(1/\beta N^2) \sum_{k,l;s}
 F(y_{kl;s})+ Y_0$ with $Y_0$ given by the
initial conditions.
Here $F(y_{kl;s})= \pm \int {\rm d}y_{kl;s} (y_{kl;s} \sqrt{W})^{-1}
= - {\rm ln}\left(2(\gamma^2 + 
2(-1)^s \tilde c_{kl;s}y_{kl;s}+ \gamma \sqrt{W})/y_{kl;s}\right)$
with $W={\gamma^2 + 4 y_{kl;s}(c_{kl;s} y_{kl;s} + (-1)^s \tilde c_{kl;s})}$
and constants  $c_{kl;s}$ and $\tilde c_{kl;s}$ given by relations
$y_{lk;s}=c_{kl;s} y_{kl;s}$ and
$x^2_{kl;s} + (-1)^s \gamma x_{kl;s} - c_{kl;s} y_{kl;s}^2 -
(-1)^s\tilde c_{kl;s} y_{kl;s} =0$.
Further, all $y_{kl;s}$ and $x_{kl;s}$ being  indicators
of the complexity of the system, $Y$ can 
be termed as the complexity parameter. 
 Now by defining $P_1=C_2 P$ with 
$C_2={\rm e}^{\int C_1 {\rm d}Y}$, the eq.(4) can be rewritten   
 as the equation governing the evolution of eigenvalues in terms of the
parameter $Y$,

\begin{eqnarray}
{\partial P_1\over\partial Y} &=&
\sum_{r=1}^2 \sum_{n=1}^N
{\partial \over \partial z_{nr}}\left[ (-1)^r \, 
 {\partial \over \partial z_{nr}}  -  (-1)^r \, \beta \, 
{\partial {\rm ln} |\Delta_N (z)| \over \partial z_{nr}}
+ \gamma z_{nr} \right] P_1
\end{eqnarray}

with $\beta=2$ and $P_1$ is related to the normalized distribution by
$\tilde P = C_2 P_1/C$.
Note  the analogy of the above equation
to that of Hermitian case \cite{ps1} but the evolution is now occuring on a
complex plane.


	The steady state of eq.(6),
$P_s\equiv |Q_N|^2 = \prod_{j< k}|\Delta_N (z)|^2
{\rm e}^{-{\gamma\over 2}\sum_{k,r} (-1)^r  \, z_{kr}^2}$,
corresponds to ${\partial P\over\partial Y} \rightarrow 0$ and 
$Y\rightarrow \infty$ which is possible when almost all 
$F(y_{kl;s})\rightarrow \infty$ or, equivalently, 
$\tilde c^2_{kl;s} = \gamma^2 c_{kl;s}$. The latter gives the condition on  
$y_{kl;s}$ and $x_{kl;s}$ for a steady state to occur.  
A choice of almost all
$y_{kl;s} \rightarrow N/(1-\tau^2)$ and
 $x_{kl;s} \rightarrow (-1)^{s-1} N \tau/(1-\tau^2)$ with $\gamma=1$ 
fulfills the condition for $\tau \rightarrow 0, \; \pm 1$ and   
can therefore lead to three different types of steady states, 
namely,  Ginibre ($\tau=0$), 
GUE ($\tau=1$) and  the ensemble of complex antisymmetric matrices or GASE 
($\tau=-1$). Here the distribution $P_s$
represents all the three cases and, in each case, agrees well
with already known distributions for GBE ($z_k$ as complex eigenvalues), 
GUE ($z_k$ as real eigenvalues) and GASE 
 (thus eigenvalues in equal and
opposite pairs).

The eq.(6) describes a transition from a given initial ensemble
(with $Y=Y_0$) to either GBE, GUE or GASE  with $Y-Y_0$ as the
transition parameter.
The nonequilibrium
states of these transitions, given by non-zero finite values of $Y-Y_0$,
are various ensembles of the complex matrices
 corresponding to varying values of
$y_{kl}$'s and $x_{kl}$'s thus modelling different complex systems.
Note the eq.(6) for $P_1\equiv P_1(\mu,Y|\mu_0,Y_0)$ has been obtained for
arbitrary initial conditions, say $P_1(\mu_0,Y_0)$;
the  distribution
$P_1(\mu,Y)= \int P_1(\mu,Y|\mu_0,Y_0) P_1(\mu_0, Y_0) {\rm d}\mu_0$
of a given NHRE can therefore be found by solving the eq.(6)
 by using a convenient initial ensemble. Just as in
the Hermitian case, the "convenience" depends on mathematical
tractability of the integrals as well as on involved physics \cite{ps1}.


   The case of non-Hermitian real matrices ($\beta=1$) can similarly
be treated. Here eigenvalues are either real or form complex
conjugate pairs and therefore if $U_n$ is an eigenvector corresponding to
the  eigenvalue $\lambda_n$, its complex conjugate $U_n^*$ 
will correspond to
the eigenvalue $\lambda_n^*$. Consider the case with $L$ real and $M$ complex
conjugate pairs of the eigenvalues with $N=L+2M$.
The rates of change of the eigenvalues and
the eigenvectors are still given by eqs. (2)
with $H_{kl;1} \equiv H_{kl}$. The distribution $P$ in this case
can be described
by $P=\int \prod_{j=1}^N f(z,z^*) g(z,z^*) \rho(H) {\rm d}H $ with
$f(z,z^*)= \prod_{j=1}^L \delta(z_j-\lambda_j)\delta_(z_j-\lambda_j^*)$,
$g(z,z^*)= \prod_{j=L+1}^{L+M} \delta(z_j-\lambda_j)
\delta_(z_j^* -\lambda_{j+M}) \delta(z_{j+M} - \lambda_j^*)
\delta(z_{j+M}^*-\lambda_{j+M}^*)$. (As obvious, here first $L$ eigenvalues
are real and rest of them complex conjugate pairs).
Proceeding similarly as for the complex case,
using eqs.(2) and equalities
${\partial fg \over \partial H_{kl}} =
- \sum_{n=1}^{L+2M} {\partial fg\over \partial  z_n}
{\partial \lambda_n \over \partial  H_{kl}}$,
${\partial^2 fg \over \partial H_{kl} H_{lk}} =
- \sum_{n=1}^{L+2M}{\partial \over \partial  z_n}\left(
{\partial fg \over \partial z_{n}}+
\sum_{m\not=n}{fg\over \lambda_m- \lambda_n}\right)$,
 one obtains 

\begin{eqnarray}
{\partial P_1\over\partial Y} &=&
 \sum_{n=1}^{L+2M}
{\partial \over \partial z_n}\left[
 {\partial\over \partial z_n}  - \beta
{\partial {\rm ln} |\Delta (z)| \over \partial z_n}  +\gamma z_n \right] P_1
\end{eqnarray}
where $Y$ is still given by the eq.(5) with $\beta=1$
($y_{kl;1} \equiv y_{kl}$
and $x_{kl;1} \equiv x_{kl}$);
 $Y= {1\over  N^2} \sum_{k,l}  F(y_{kl})$
with $F(y_{kl})=  \int {\rm d}y_{kl} (y_{kl} \sqrt{W})^{-1}
= - {\rm ln}
\left[ 2(\gamma-2\tilde c_{kl}y_{kl}+\gamma \sqrt{W})/y_{kl}\right]$,
 $W= \gamma^2 + 4 y_{kl} (c_{kl} y_{kl} - \tilde c_{kl})$ and
 $c_{kl}$, $\tilde c_{kl}$ given by relations
$y_{lk}=c_{kl} y_{kl}$ and
$x^2_{kl} - \gamma x_{kl} - c_{kl} y_{kl}^2 + \tilde c_{kl} y_{kl} =0$.
 The steady state again occurs for $Y \rightarrow \infty$ 
with corresponding solution of eq.(7) is given by  
$P_1(Y\rightarrow \infty)= P_s = |\Delta_N(z)|
\left[ \prod_{i=1}^N  {\rm e}^{- \gamma z_i^2}
{\rm erfc}(z_i-z_i^*)\right]^{1/2}$. As before, the limit 
$Y \rightarrow \infty$ can be obtained for 
$y_{kl} \rightarrow N/(1-\tau^2)$, 
 $x_{kl} \rightarrow N \tau/(1-\tau^2)$ and $\gamma=1$ 
with $\tau \rightarrow 0, \; \pm 1$; the steady state is an 
ensemble of real matrices ($\tau=0$),  
a GOE ($\tau=1$) and  
a Gaussian ensemble of real anti-symmetric matrices ($\tau=-1$).    
The distribution $P(z;\tau=0,\pm 1)$ is in agreement with the results
obtained in \cite{nh}, for the corresponding ensembles, by  a 
different method.

	As in the Hermitian case \cite{ap,ps1}, a direct integration of  
eq.(6) (over all $z_i$, $i=n+1\rightarrow N$) can be used to obtain the 
 $n^{\rm th}$ order density correlator $R_n (z_1,..z_n; Y)$, defined by
 $R_n = { N! \over {(N-n)!}}\int P(z, Y)
{\rm d}^2z_{n+1}..{\rm d}^2z_N$
with ${\rm d}^2 z_n\equiv {\rm d}z_{n1} {\rm d}z_{n2}$.
%
%
However	for real applications, it is important to consider the
limit $N\rightarrow \infty$ for fixed $n$. To take the limit 
meaningfully for $R_1(z_1)$, 
a change of variable $z_1 \rightarrow e$ is required  
where $z_1={\sqrt N} e$ (as $R_1(z_1) \approx O(N)$). An integration 
of eq.(6), followed by the change $z_1\rightarrow e$, will give the
evolution equation of $R_1(e)$: 
\begin{eqnarray}
{\partial R_1\over\partial Y} =
  \sum_{r=1}^2 {\partial \over \partial e_r}\left( \gamma \, e_r -
 2 \,  (-1)^r \, {\bf P}\int {\rm d}^2e' R_1 (e')  {e_r - e'_r\over |e- e'|^2}
\right) R_1\nonumber
\end{eqnarray}
(${\bf P}$ as the principle part of the integral).
Here the terms containing $2^{\rm nd}$ order cluster functions 
 and the diffusion term (see \cite{ap}) have been neglected, 
both being $O(N)$ (or more)  smaller than other terms.
%
%
    For $n>1$, the correlators should be
unfolded before taking the limit $N\rightarrow \infty$:
$R_n(\zeta_1,..,\zeta_n;\Lambda)={\rm Lim} N\rightarrow \infty \;
{{\it R}_n(z_1,..,z_n;Y) \over
{\it R}_1(z_1;Y)...{\it R}_1(z_n;Y)}$
with $\zeta=\int^{\zeta} {\it R}_1^{1/2}(z;Y){\rm d}z$ 
and $\Lambda=(Y-Y_0) R_1$. 
Again ,first, the eq.(6) is integrated to get the  hierarchic relations 
among $R_n(z)$ and then the limit $N\rightarrow \infty$ is applied after 
replacing $R_n(z)$ by $R_n(\zeta)$. This gives
\begin{eqnarray}
{\partial R_n \over\partial \Lambda} &=&  \sum_{r=1}^b \sum_{j=1}^n (-1)^r  \,
 {\partial \over \partial \zeta_{jr}} \left[ |\Delta_n|^\beta \, 
 {\partial \over \partial \zeta_{jr}} 
{R_n \over |\Delta_n|^\beta }- \beta I(z_j)\right]
\end{eqnarray}
with $I(\zeta_j)\equiv \int_{-\infty}^{\infty} {\rm d} \zeta_{n+1} R_{n+1}
{\partial {\rm ln} |\zeta_{j}-\zeta_{n+1}|/ \partial \zeta_{jr}}$  
 and 
$b=2$ (for simplification, $\gamma$ is chosen to be unity).
Here the rescaling of parameter $Y$ was 
required to see the smooth transition in $R_n$ (the evolution in 
$R_n(\zeta)$  takes place for finite values of $Y R_1$).
%
%
The eq.(8) is obtained by neglecting the linear 
drift of the eigenvalues which is dominated, by a factor $R_1$, by their 
diffusion and mutual repulsion. In fact, the linear restoring
force, responsible for the global behaviour of the density of levels,
is entirely negligible on scales at which local fluctuations occur. On the
other hand, the diffusion is ineffective on the global scale 
(see equation for $R_1$).
As discussed above, the transition for $R_n$ occurs
on the scales determined by $Y \approx R_1^{-1}$,
while, for $R_1$, the corresponding
scale is given by $Y \approx N\; R_1^{-1}$. 
This indicates a clear separation of
the scales of the global and local behaviour of the density. 
The hierarchical equation of correlations for the
real assymetric case can be obtained by integrating eq.(7) which
will again lead to a relation similar to eq.(8) but now 
$\beta=1$. Further, for all those $\zeta_j$ which correspond to 
real eigenvalues,
 the $\sum_r$ is to be dropped with $\zeta_{jr}$ replaced by $\zeta_j$.




	An alternative route to obtain correlations is by exploiting  the
 connection of eq.(6) and eq.(7) to CS Hamiltonian. 
This can be shown by using the transformation $\Psi=P/|Q_N|^{\beta/2}$ 
in eq.(6) (and eq.(7)) reducing it in a form
 ${\partial \Psi \over\partial Y} = \hat H \Psi$
where the 'Hamiltonian'  $\hat H = {1\over 2} \left({\mathcal H} + {\mathcal H}^* \right)$ turns out to be a variant of the
CS Hamiltonian in two dimensions (for simplification take $\gamma=1$):
\begin{eqnarray}
{\mathcal H} =\sum_i  {\partial^2 \over \partial r_i^2} -
g \sum_{i,j; i<j} {1\over r^2_{ij}}
-G \sum_{i,j,k; i,j\not=k} \,  q_{ij} \, 
 {{\bf r_{ki}}.{\bf r_{kj}} \over r^2_{ki} r^2_{kj}} -
 {1\over 4} \, \sum_i  r_i^2 \nonumber
\end{eqnarray}
with ${\bf r}_i\equiv z_i$, ${\bf r}_{ki}\equiv z_k-z_i$ and
$ r_{ki} \equiv |{\bf r}_{ki}|$ and  $q_{ij} = \beta - 4 \delta_{ij}$. Here $g=0$, $G={\beta \over 4}$ (with $\beta=1$ for real NHRM case and $\beta=2$ for the complex NHRM) and, analogous to the complex
Hermitian case ($G=0, g=0$), the inverse square term drops out
for the complex non-Hermitian case too. Further here the particles are fermions (as 
$g \not= G$) similar to the Hermitian case.  
As $Y\rightarrow \infty$, the particles are in their ground state
 $\psi_0 =  \prod_{j< k}^N |{\bf r_i}- {\bf r_j}|^\beta \,
{\rm e}^{-{1\over 4}\sum_k (r_k^2 + r_k^{*2})}$ 
 with a distribution $\psi_0^2$; note $\psi_0$ gives the correct 
form for $P(Y\rightarrow \infty)$. 
%
%
The eigenstates and the eigenvalues
of the Hamiltonian $\hat H$ for the case $g=0$ are well known \cite{ak}:
The "state" $\psi$ or $P(\mu,Y|\mu_0,Y_0)$ can then formally
be expressed as a
sum over the  eigenvalues and eigenfunctions  which on integration over  the
initial state $P(\mu_0,Y_0)$  would
lead to the joint probability distribution $P(\mu,Y)$ and
thereby static (at a single parameter value) density correlations $R_n$. The
above correspondence can also be used to map the multi-parametric
correlations  of levels to multi-time correlations of
 the particle-positions \cite{ps1}.  Although the explicit calculations
of correlations, involves technical handling of  various integrals and
is still an unfinished task, nontheless our study reveals an important
connection. The
level correlations of different complex systems need not be studied
separately, a thorough probing of the particle correlations of
CS type Hamiltonian will give all the required information. The CS system being
integrable in nature, the semiclassical
techniques can also be very successful for the probing.

	 The reasons for the correspondence between Gaussian NHRM and CS
Hamiltonian are worth paying attention. Note the analogy with
harmonic oscillator type confining potential in the CS system
results from the Gaussian nature of the
ensemble. The  correspondence with $1/r^2$ term
comes from the mutual repulsion between eigenvalues. The  mathematical
origin of the latter lies in the transformation from matrix 
space to eigenvalue-eigenvector
space which is same for all the non-hermitian ensembles
(belonging to same symmetry class, irrespective of matrix element distribution).
It should be possible, therefore,
to map the non-Gaussian NHRM also to a variant of CS Hamiltonian,
although with a different type of confining potential. For
$\rho(H) \propto {\rm e}^{-f(H)}$, the correspondence can be shown by
following the similar steps as used for the Hermitian case \cite{ps1}.
	Also note that
the  short range correlations of all the NHRMs are dominated by 
 the mutual repulsion, and therefore are expected to be
nearly similar within the same symmetry class.


	In this paper, we have studied the statistical properties of the
 eigenvalues of non-Hermitian systems.
 We find that the distribution of the eigenvalues is governed
by a diffusion equation in which system-dependence enters only through the
evolution  parameter
$ \Lambda \propto Y-Y_0$ related to complexity of the system. 
The  spectral  
correlators for a given NHRE also depend only on the associated $\Lambda$ 
value. 
It is possible that
widely different systems with different values of the distribution
parameters may share a same $\Lambda$-value. Such systems will thus
have similar statistical featurs.
Furthermore as the eigenvalue distribution for each complex
system appears as a general state of CS system, any two such states,
for example $\psi(Y_1)$ and $\psi(Y_2)$, being
related by "time" evolution operator $U(Y_2, Y_1)$, the eigenvalue
distributions of the complex systems corresponding to $Y_1$ and $Y_2$
will also be connected. This would also reflect in  their physical
properties based on spectral fluctuations e.g conductance (assuming
existence of ergodicity, that is, ensemble averages same as spectral
averages).

The apperance of CS Hamiltonain is not
restricted only to the spectral properties; it has been known to
manifest itself in other properties of complex systems too \cite{cpm}.
A detailed investigation of
CS hamiltonian in arbitrary dimension can therefore give a lot of
useful information about variety of complex systems and is very much
desirable.
.....

\end{document}